\documentclass[11pt,a4paper]{article}
\usepackage[latin9]{inputenc}
\usepackage{amsmath}
\usepackage{amsthm}
\usepackage{amssymb}
\usepackage{graphicx}
\usepackage{wasysym}

\makeatletter

\usepackage{amsfonts}
\usepackage{amsthm}
\usepackage{epstopdf}

\newif\iffigs\figstrue

\makeatother

\begin{document}
\begin{center}
\textsc{\Large{}{}\ \\[0pt] \vspace{5mm}
 Instantons with Quantum Core}{\Large{}{} }\\[0pt] 
\par\end{center}

\begin{center}
\vspace{35pt}
 \textsc{V. F. Mukhanov$^{~a,b}$, A. S. Sorin$^{~c,d,e}$}\\[15pt] 
\par\end{center}

\begin{center}
{$^{a}$ Ludwig Maxmillian University, \\[0pt] Theresienstr. 37,
80333 Munich, Germany\\[0pt] }e-mail: \textit{\small{}{}mukhanov@physik.lmu.de}{\small{}\vspace{10pt}
 }{\small\par}
\par\end{center}

\begin{center}
{$^{b}$ {\small{}{}Korea Institute for Advanced Study\\[0pt] Seoul,
02455, \ Korea\\[0pt] }}e-mail: \textit{\small{}{}mukhanov@physik.lmu.de}{\small{}\vspace{10pt}
 }{\small\par}
\par\end{center}

\begin{center}
{$^{c}$ {\small{}{}Bogoliubov Laboratory of Theoretical Physics\\[0pt]
Joint Institute for Nuclear Research \\[0pt] 141980 Dubna, Moscow
Region, Russia \\[0pt] }}e-mail: \textit{\small{}{}sorin@theor.jinr.ru}{\small{}\vspace{10pt}
 }{\small\par}
\par\end{center}

\begin{center}
{$^{d}$ {\small{}{}National Research Nuclear University MEPhI\\[0pt]
(Moscow Engineering Physics Institute),\\[0pt] Kashirskoe Shosse
31, 115409 Moscow, Russia}}\vspace{10pt}
 
\par\end{center}

\begin{center}
{$^{e}$ {\small{}{}Dubna State University, \\[0pt] 141980 Dubna
(Moscow region), Russia}}\vspace{10pt}
 
\par\end{center}

\begin{center}
\vspace{3mm}
 \textbf{{Abstract} } 
\par\end{center}

We consider new instantons that appear as a result of accounting
for quantum fluctuations. These fluctuations naturally regularize
the $O(4)$ singular solutions abandoned in Coleman's theory. In the previous
works \cite{MRS1,MRS2} we showed how new instantons modify
the widely accepted picture of false vacuum decay in two particular
examples of exactly solvable potentials. Here we generalize our consideration
to arbitrary potentials and provide a general theory of these new
instantons with quantum cores in which vacuum fluctuations dominate.
We develop a method that allows us to determine the parameters of
instantons for generic potentials not only in the thin-wall approximation
but also in the cases where this approximation fails. Unlike the Coleman
instantons, the instantons with quantum cores always exist in the
cases where the vacuum must be unstable. 

\newpage{}

\section{Introduction}

\begin{figure}[hbt]
\begin{centering}
\includegraphics[height=80mm]{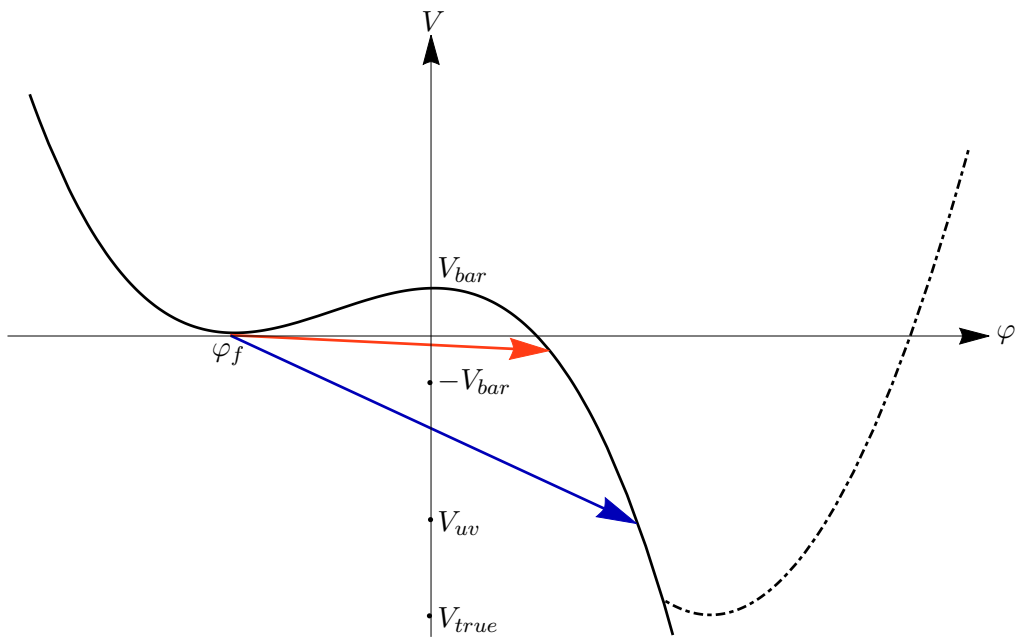} 
\par\end{centering}
\vspace*{-4.2cm}
 \hspace*{4.1cm} $\varphi_{f}$

\vspace*{-4.8cm}
 \hspace*{6.99cm}$V$

\vspace*{3.6cm}
 \hspace*{14.55cm}$\varphi$

\vspace*{0.2cm}
 \hspace*{7.2cm}$-V_{bar}$

\vspace*{1.4cm}
 \hspace*{7.2cm}$V_{uv}$

\vspace*{0.8cm}
 \hspace*{7.2cm}$V_{true}$

\vspace*{-5.12cm}
 \hspace*{7.2cm}$V_{bar}$

\vspace{1.5cm}

\vspace*{3.3cm}
\caption{The potential with a metastable vacuum at $\varphi_{f}$, which must
decay via instantons, regardless of whether this potential is unbounded
from below or has the second true minimum (represented by the dot-dashed
line). For each concrete potential, there is a full spectrum of
instantons. Those instantons corresponding to the transitions represented
by the upper red line describe the thin-wall bubbles for which the
potential in the center of the emerging bubble satisfies the condition
$\left|V_{uv}\right|\ll V_{bar}$. The instantons corresponding to
the deep subbarrier transition (as shown, for example, by the lower blue
line) are dominated by ``friction'' and cannot be described in the
thin-wall approximation.}
\label{Figure1} 
\end{figure}

Let us consider a real scalar field with the potential $V\left(\varphi\right)$
shown in Fig.\ref{Figure1}. For simplicity, we normalize this potential to zero
at the local minimum at $\varphi_{f}<0$, i.e., $V\left(\varphi_{f}\right)=0$.
The potential has a maximum at $\varphi=0$, where $V\text{\ensuremath{\left(0\right)}}=V_{bar}$
is the hight of the barrier. If at $\varphi>0$ the potential is unbounded
from below or negative at the second minimum, which corresponds to the
true vacuum, then the false vacuum at $\varphi_{f}$ is obviously
metastable and must decay. As a result of the sub-barrier tunneling,
the critical bubbles are formed, which are filled with a new phase $\varphi>0$.
They expand and eventually collide, filling the space with a new phase.

\vspace{0.5cm}


At first sight, the problem looks very similar to the problem of tunneling
the ``particle'' with $N$ degrees of freedom, characterized by
generalized coordinates $\mathbf{q\mathnormal{\equiv(}}q_{1},...,q_{N})$,
from the local minimum of the potential $V\left(\mathbf{q}\right)$.
However, this analogy is incomplete and, moreover, even confusing.
In fact, for a system with a finite number degrees of freedom, the subbarrier
transition is reversible, i.e., the system can return to the original
minimum after tunneling. In field theory, vacuum decay is irreversible.
Although there are solutions that seem to describe a transition
that takes us back to a false vacuum state, they correspond to contracting
bubbles with very improbable initial conditions of zero measure. Moreover,
we would like to emphasize that the scalar field potential $V\left(\varphi\right)$
should in no way be considered as an analogue of the potential $V\left(\mathbf{q}\right)$.
To make this clear, it is convenient to represent the action for the
scalar field 
\begin{equation}
S=\int\left(\frac{1}{2}\eta^{\alpha\beta}\varphi_{,\alpha}\varphi_{,\beta}-V\left(\varphi\right)\right)d^{4}x\,,\label{1a}
\end{equation}
where $\eta^{\alpha\beta}$ is the Minkowski metric, in the following equivalent
form:
\begin{equation}
S=\int(\mathcal{K-V)}dt\,,\label{1}
\end{equation}
where 
\begin{equation}
\mathcal{K}\equiv\frac{1}{2}\int\left(\partial_{t}\varphi_{\mathbf{x}}\right)d^{3}x\,,\label{2}
\end{equation}
and 
\begin{equation}
\mathcal{V}\equiv\int\left(\frac{1}{2}\left(\partial_{i}\varphi_{\mathbf{x}}\right)^{2}+V\left(\varphi_{\mathbf{x}}\right)\right)d^{3}x\,.\label{3}
\end{equation}
From here it is obvious that we consider a system with an infinite
number of degrees of freedom, described by generalized coordinates
$\varphi_{\mathbf{x}}(t)\equiv\varphi\left(\mathbf{x\mathnormal{,t}}\right)$,
which characterizes the strength of the field at each point in space,
and the spatial coordinates $\mathbf{x=}\left(x^{1},x^{2},x^{3}\right)$
simply enumerate the degrees of freedom. Accordingly, $\mathcal{K}$
and $\mathcal{V}$ are the kinetic and potential energies of the system.
Therefore, the functional $\mathcal{V}\left(\varphi_{\mathbf{x}}\right)$
(\textit{but not} $V\left(\varphi\right)$) plays the role of the potential
when we consider sub-barrier tunneling in quantum field theory. If
the minimum at $\varphi_{f}$ corresponds to a false vacuum, then
the potential $\mathcal{V}\left(\varphi_{\mathbf{x}}\right)$ is always
unbounded from below even for the bound potential $V\left(\varphi\right)$
with the second true minimum of depth $V\left(\varphi_{tr}\right)=-\epsilon$.
Indeed, in this case, the constant field configuration $\varphi=\varphi_{tr}$
has the potential energy $\mathcal{V}\left(\varphi_{\mathbf{x}}=\varphi_{tr}\right)=-\epsilon\times volume$
and tends to minus infinity as the volume grows. Therefore, the false
vacuum decay is always analogous to the sub-barrier escape of the
``particle with an infinite number degrees of freedom'' from a local
well in the unbounded potential. To describe the sub-barrier transition,
it is convenient to perform the Wick rotation and switch from Minkowski
to Euclidean time $\tau=\mathsf{i}\, t$. The action (\ref{1}) then becomes $S=\mathsf{i} \,S_{E}$,
where 
\begin{equation}
S_{E}=\int\left(\frac{1}{2}\left(\partial_{\tau}\varphi\right)^{2}+\frac{1}{2}\left(\partial_{i}\varphi\right)^{2}+V\left(\varphi\right)\right)d^{3}xd\tau\label{4a}
\end{equation}
is the Euclidean action. Assuming that the sub-barrier transition
from the false vacuum occurs in the ``time interval'' $0\geq\tau>-\infty$,
we must find field configurations with $\varphi\left(\tau\rightarrow-\infty,\mathbf{x}\right)=\varphi_{f}$,
matching the classically allowed state $\varphi\left(\tau=0,\mathbf{x}\right)$
with $\partial\varphi/\partial\tau=0$ and $\mathcal{V}\left(\varphi\left(\mathbf{x}\right)\right)=0$.
It is clear that the main contribution to the tunneling rate, proportional
to $\exp\left(iS\right)=\exp\left(-S_{E}\right)$, should come from
those field configurations that minimize the Euclidean action, i.e.,
satisfy the equation 
\begin{equation}
\partial_{\tau}^{2}\varphi+\Delta\varphi-V'=0\,,\label{5a}
\end{equation}
where $V'\equiv dV/d\varphi.$ Therefore, to calculate the tunneling rate,
one must first find the corresponding solutions of equation (\ref{5a}).

\section{The Coleman Instanton and Quantum Fluctuations}

\begin{figure}[hbt]
\begin{centering}
\includegraphics[height=80mm]{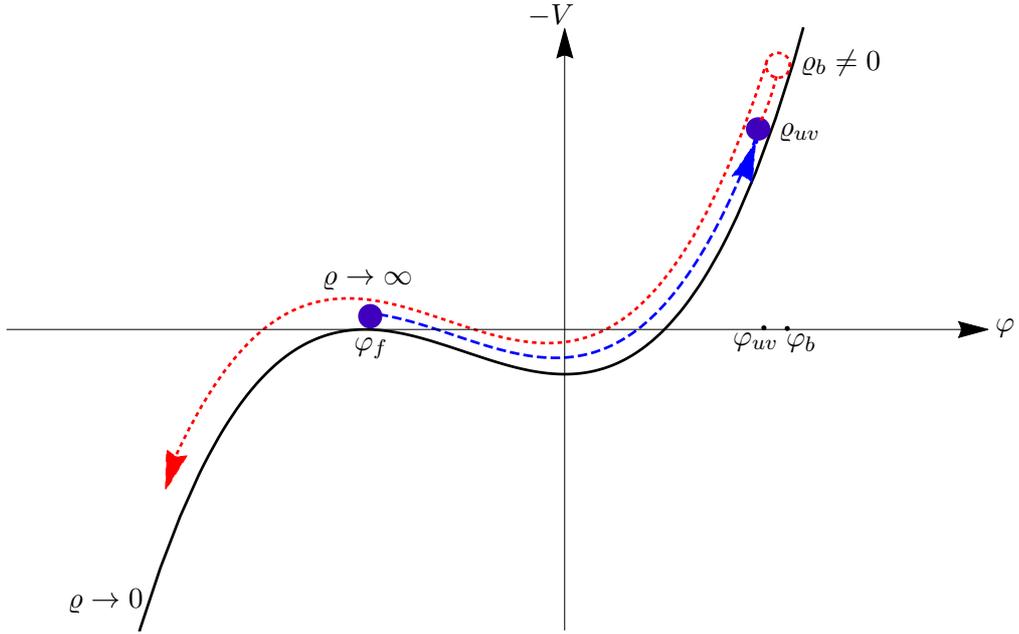} 
\par\end{centering}
\vspace*{-4.2cm}
 \hspace*{5.99cm} $\varphi_{f}$

\vspace*{-4.8cm}
 \hspace*{8.4cm}$-V$

\vspace*{3.6cm}
 \hspace*{14.55cm}$\varphi$

\vspace*{-0.3cm}
 \hspace*{11.1cm}$\varphi_{uv}$

\vspace*{-0.46cm}
 \hspace*{11.8cm}$\varphi_{b}$

\vspace*{-4.15cm}
 \hspace*{11.99cm}$\varrho_{b}\neq0$

\vspace*{0.4cm}
 \hspace*{11.7cm}$\varrho_{uv}$

\vspace*{1.5cm}
 \hspace*{5.7cm}$\varrho\rightarrow\infty$

\vspace*{3.8cm}
 \hspace*{2.35cm}$\varrho\rightarrow0$

\vspace*{0.5cm}
\caption{The solution of equation (\ref{5a-1}) for the instanton is mathematically
equivalent to the solution of the equation for the ``particle''
with friction in the inverted potential $-V$. It starts its motion
at $\varrho\rightarrow\infty$ on the top of the inverted potential
and moves toward positive $\varphi$ when the ``time'' $\varrho$
decreases, as shown by the dashed blue line. At some $\varrho_{b}\protect\neq0$,
the ``velocity'' $\dot{\varphi}$ vanishes and the ``particle''
bounces off and starts moving in the direction of negative $\varphi$,
as shown by the dotted red line. Only when $\varrho_{b}=0$, the motion
ends at the highest point and does not go further back. This is the
case of the nonsingular Coleman instanton, which at this moment materializes
as a bubble in the Minkowski space. For any finite $\varrho_{b}$ the
bouncing classical trajectories are singular. However, as we have
shown, the classical solutions corresponding to these singular trajectories
can be trusted only up to $\varrho_{uv}>\varrho_{b}$ because at this
moment quantum fluctuations begin to dominate and therefore the part
of the solution indicated by the dotted red line must be ignored and
only the part, shown by the dashed blue line, can be trusted. At $\varrho=\varrho_{uv}$,
the bubble with the quantum core of size $\varrho_{uv}$, where the
potential is shifted to $V_{uv}$, emerges from under the barrier
and materializes in the Minkowski space.}
\label{Figure2} 
\end{figure}

From general considerations, one can expect that among all possible solutions
of equation (\ref{5a}), the solution with the highest possible
symmetry has the smallest action and dominates the decay rate \cite{Coleman,CGM}.
Therefore, Coleman proposed to consider $O(4)$ - invariant
solutions for which $\varphi$ depends only on $\varrho=\sqrt{\tau^{2}+\mathbf{x}^{2}}$,
that is, $\varphi\left(\tau,\mathbf{x}\right)=\varphi\left(\varrho\right)$.
In this case, equation (\ref{5a}) is simplified to the ordinary
differential equation 
\begin{equation}
\ddot{\varphi}(\varrho)+\frac{3}{\varrho}\,\dot{\varphi}(\varrho)-V'=0\,,\label{5a-1}
\end{equation}
where the dot denotes the derivative with respect to $\varrho.$ This
reduces the problem to studying ``the motion of a particle with one
degree of freedom'' in the presence of friction in the inverted potential
$V$ (see Fig.\ref{Figure2}) \footnote{Note that the friction term in (\ref{5a-1}) comes from $\Delta\varphi$ in (\ref{5a}).}.

\vspace{0.6cm}


In order to find a particular solution of (\ref{5a-1}) that makes
a dominant contribution to the tunneling rate, we need to decide what
are the proper initial or, alternatively, boundary conditions for
this solution. One of them is obvious: namely, if the field is ``initially''
in the false vacuum state at $\tau\rightarrow-\infty$, then 
\begin{equation}
\varphi\left(\varrho\rightarrow\infty\right)=\varphi_{f}\,.\label{6a}
\end{equation}
Assuming that at the ``moment of emergence'' at $\tau=0$ the field
in the center of the bubble is equal to $\varphi\left(\varrho=0\right)=\varphi_{c}$,
we can approximate (\ref{5a-1}) near this point, i.e., for $\left|\varphi-\varphi_{c}\right|\ll\varphi_{c}$,
as 
\begin{equation}
\ddot{\varphi}(\varrho)+\frac{3}{\varrho}\,\dot{\varphi}(\varrho)-V_{c}'\simeq0\,,\label{7a}
\end{equation}
where $V_{c}'=V'\left(\varphi_{c}\right)$. This equation has the
following general solution: 
\begin{equation}
\varphi\left(\varrho\right)=\varphi_{c}+\frac{V_{c}'}{8}\,\varrho^{2}+\frac{C}{\varrho^{2}}\,,\label{8a}
\end{equation}
where the integration constant $C$ must be taken equal to zero in order to
be in agreement with $\varphi\left(\varrho=0\right)=\varphi_{c}$.
Therefore, 
\begin{equation}
\dot{\varphi}(\varrho=0)=0\label{9a-1}
\end{equation}
is the only possible choice for the boundary condition at $\varrho=0$
that does not lead to a singularity. The $O(4)$-solution
of equation (\ref{5a-1}) with boundary conditions (\ref{6a}) and
(\ref{9a-1}), if it exists, is unique and is called the Coleman instanton.

\begin{figure}[hbt]
\begin{centering}
\includegraphics[height=80mm]{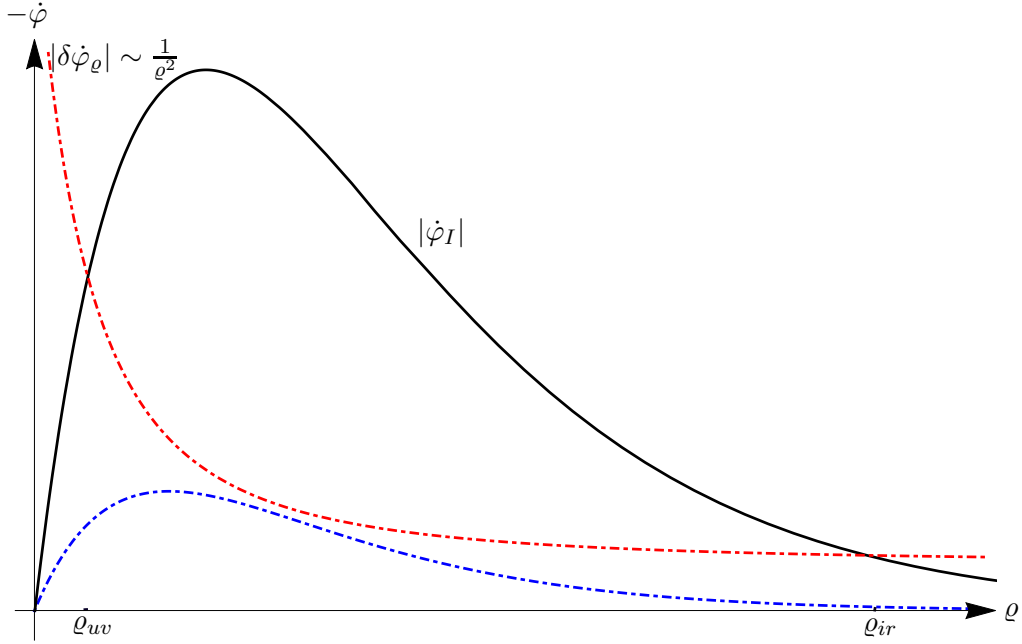} 
\par\end{centering}
\vspace*{-8.7cm}
 \hspace*{1.3cm} $-\dot{\varphi}$

\vspace*{7.4cm}
 \hspace*{14.55cm}$\varrho$

\vspace*{-0.28cm}
 \hspace*{12.7cm}$\varrho_{ir}$

\vspace*{-0.5cm}
 \hspace*{2.3cm}$\varrho_{uv}$

\vspace*{-5.64cm}
 \hspace*{6.84cm}$|\dot{\varphi}_{I}|$

\vspace*{-2.8cm}
 \hspace*{2cm}$|\delta\dot{\varphi}_{\varrho}|\sim\frac{1}{\varrho^{2}}$

\vspace{1.5cm}

\vspace*{6.5cm}
\caption{The red dot-dashed line indicates the level of quantum fluctuations. The
classical solution with the Coleman boundary conditions, represented
by the blue dot-dashed line, is completely hidden under the quantum fluctuations
and is not trustworthy. For the Coleman instanton, shown by the black
solid line, the reliable part of the classical solution lies in the
region $\varrho_{uv}<\varrho<\varrho_{ir}$, where the strength of
the field exceeds the level of the quantum fluctuations.}
\label{Figure3} 
\end{figure}

We showed in previous publications that in some cases such instantons
either do not exist \cite{MRS2} or they lead to an unexpectedly
large vacuum decay rate \cite{MRS1}. As we demonstrated in \cite{MRS2,MRS1},
both these problems can be resolved by involving unavoidable
quantum fluctuations. These fluctuations can either regularize the
solutions, which were previously singular, or completely saturate
the instantons leading to too large a decay rate. In fact, the classical
solution can be trusted only if the field strength exceeds the level
of vacuum fluctuations at the appropriate scale. In the case of a
massless scalar field, the ``typical'' amplitude of quantum
fluctuations and their time derivative in scales $r$ are approximately
equal to (see, e.g., \cite{Mukhanov,MW}) 
\begin{equation}
\left\vert \delta\varphi_{r}\right\vert \simeq\frac{\sigma}{r}\,,\quad\left\vert \delta\dot{\varphi}_{r}\right\vert \simeq\frac{\sigma}{r^{2}}\,,\label{10}
\end{equation}
where $\sigma$ is the number of the order of unity\footnote{We use the Planck units where all fundamental constants are set equal
to one.}. The massless scalar field is a shift invariant, and to determine
when quantum fluctuations begin to saturate the classical field, we
must compare either the typical change of this field on scales $r$
or its time derivative with the amplitude given in (\ref{10}). Both
equations lead to the same result (up to a numerical coefficient of the order
of unity). Therefore, to be concrete, we compare the time derivative for
the instanton solution $\dot{\varphi_{I}}\left(\varrho\right)$ with
$\left\vert \delta\dot{\varphi}_{r}\right\vert $. It is clear that
the instanton solution is trustworthy only at such $r$ for which
\begin{equation}
\dot{\varphi_{I}}\left(\sqrt{\tau^{2}+r^{2}}\right)>\frac{\sigma}{r^{2}}\,.\label{11}
\end{equation}
In Fig.\ref{Figure3},  we have plotted two solutions for the Coleman instanton versus
the level of quantum fluctuations at $\tau=0$ when $r=\varrho.$
The instanton lying entirely under the $\sigma/\varrho^{2}$ curve
is never reliable, while the other is reliable only in the $\varrho_{uv}<\varrho<\varrho_{ir}$
region, where both the ultraviolet and the  infrared cutoff scales,
$\varrho_{uv}$ and $\varrho_{ir}$, are two different solutions of the following
equation: 
\begin{equation}
\dot{\varphi}_{I}\left(\varrho\right)\simeq\frac{\sigma}{\varrho^{2}}\,.\label{12}
\end{equation}



It follows that Coleman's boundary condition (\ref{9a-1}) formulated
in the deep ultraviolet region is never reliable. In turn, the boundary
condition formulated at a small finite $\varrho_{b}\neq0$ does not
a priori exclude solutions with the non-vanishing integration constant
$C$ in (\ref{8a}). A possible singularity that can occur in this
case at $\varrho=0$ is anyway regularised by the ultraviolet cutoff.
This leads to the appearance of the whole class of new instantons that
can be parametrized by $\varrho_{b}$.

\section{New Instantons}

To determine the plausible boundary condition at $\varrho_{b}$, we
first calculate the potential energy (\ref{3}) at $-\varrho_{b}\leq\tau\leq 0$.
For the $O\left(4\right)$-invariant solution $\varphi\left(\varrho\right)$,
the integral in (\ref{3}) is simplified to 
\begin{equation}
\mathcal{V}\left(\tau\right)=4\pi\int_{\varrho_{b}}^{\infty}\left[\frac{1}{2}\left(1-\frac{\tau^{2}}{\varrho^{2}}\right)^{3/2}\dot{\varphi}^{2}+\left(1-\frac{\tau^{2}}{\varrho^{2}}\right)^{1/2}V\right]\varrho^{2}d\varrho+\mathcal{V}_{r_{b}}\left(\tau\right)\,,\label{13}
\end{equation}
where $\mathcal{V}_{r_{b}}\left(\tau\right)$ is the contribution
to the potential energy from the core of the bubble of size $r_{b}\left(\tau\right)=\sqrt{\varrho_{b}^{2}-\tau^{2}}$. Integrating by parts and using equation of motion (\ref{5a-1}),
we can rewrite the second term under the integral as 
\begin{align}
 & 4\pi\int_{\varrho_{b}}^{\infty}\left(1-\frac{\tau^{2}}{\varrho^{2}}\right)^{1/2}V\varrho^{2}d\varrho=\frac{4\pi}{3}\int_{\varrho_{b}}^{\infty}Vd\left(\varrho^{2}-\tau^{2}\right)^{3/2}\nonumber \\
 & =\frac{4\pi}{3}\left(\left.\left(\varrho^{2}-\tau^{2}\right)^{3/2}V\right|_{\varrho_{b}}^{\infty}-\int_{\varrho_{b}}^{\infty}\left(\varrho^{2}-\tau^{2}\right)^{3/2}\left(\ddot{\varphi}+\frac{3}{\varrho}\dot{\varphi}\right)\dot{\varphi}d\varrho\right)\,.\label{14}
\end{align}
If we remove here the term with the second derivative by further integration
by parts and insert the result in (\ref{13}), we obtain 
\begin{equation}
\mathcal{V}\left(\tau\right)=2\,\pi\,\tau^{2}\int_{\varrho_{b}}^{\infty}\sqrt{1-\frac{\tau^{2}}{\varrho^{2}}}\,\dot{\varphi}^{2}d\varrho+\frac{2\,\pi}{3}\,\dot{\varphi}_{b}^{2}\,r_{b}^{3}-\frac{4\,\pi}{3}\,V_{b}\,r_{b}^{3}+\mathcal{V}_{r_{b}}\left(\tau\right)\,,\label{15}
\end{equation}
where $\dot{\varphi}_{b}\equiv\dot{\varphi}\left(\varrho_{b}\right)$,
$V_{b}\equiv V\left(\varphi\left(\varrho_{b}\right)\right)$ and we
have assumed that $\dot{\varphi}^{2}\varrho^{3}$ and $V\varrho^{3}$
vanish as $\varrho\rightarrow\infty$. Since we have normalized the
potential $V$ in the false vacuum to zero, the total energy of the
bubble must be zero. Taking into account that $\partial_{\tau}\varphi=\frac{\tau\,\dot{\varphi}}{\sqrt{\tau^2+\mathbf{x}^2}}$,
we find that the kinetic energy (\ref{2}) vanishes at $\tau=0$ and
therefore the bubble can materialize or, in other words, emerge from
under the barrier only when
\begin{equation}
\mathcal{V}\left(0\right)=\frac{2\,\pi}{3}\,\dot{\varphi}_{b}^{2}\,\varrho_{b}^{3}-\frac{4\,\pi}{3}\,V_{b}\,\varrho_{b}^{3}+\mathcal{V}_{\varrho_{b}}(0)\label{16}
\end{equation}
vanishes. It is quite natural to assume (see the next Section for a more
detailed justification), that the total energy of the central part
of the bubble with the radius $r_{b}=\varrho_{b}$ is mainly due to the
shift of the potential energy density $\mathcal{V}_{\varrho_{b}}(0)$ to $V_{b}$ and is therefore
equal to 
\begin{equation}
\mathcal{V}_{\varrho_{b}}(0)=\frac{4\,\pi}{3}\,V_{b}\,\varrho_{b}^{3}\,,\label{17}
\end{equation}
compensating the second term in (\ref{16}). Since the potential energy
$\mathcal{V}\left(0\right)$ must vanish, we conclude that the only
possible boundary condition we can impose at $\varrho_{b}$ is 
\begin{equation}
\dot{\varphi}\left(\varrho_{b}\right)=0\,.\label{18}
\end{equation}
The solutions of equation (\ref{5a-1}) with the boundary conditions
(\ref{6a}) and (\ref{18}) give us the whole class of new instantons
parametrized by $\varrho_{b}$ at which the corresponding classical
solution bounces.

\section{Quantum Core}

\begin{figure}[hbt]
\begin{centering}
\includegraphics[height=80mm]{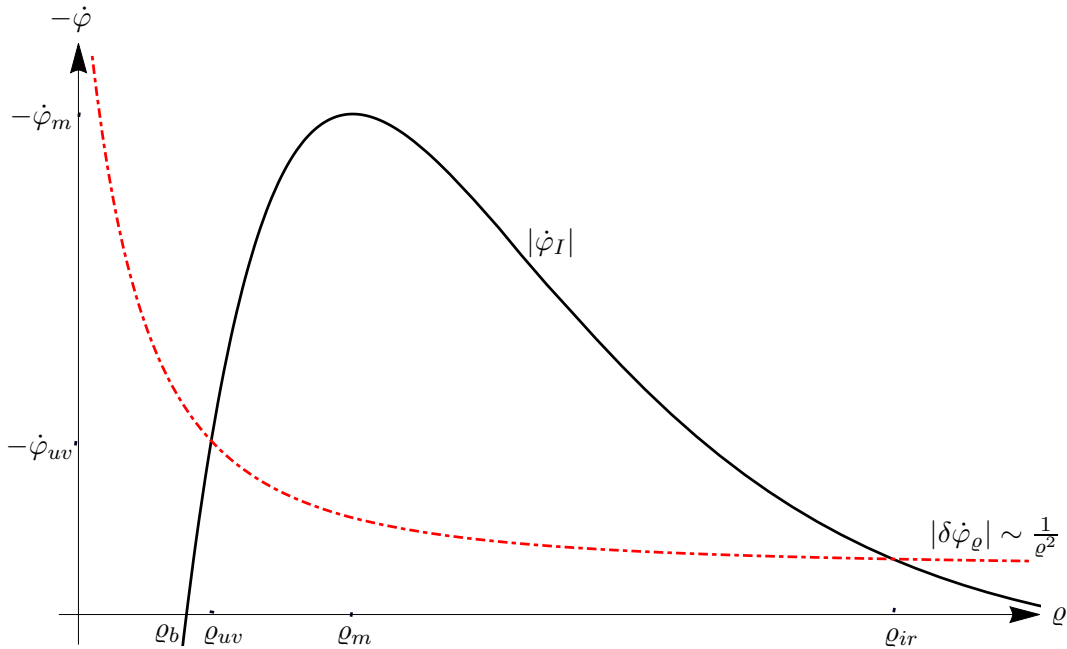} 
\par\end{centering}
\vspace*{-8.7cm}
 \hspace*{1.3cm} $-\dot{\varphi}$

\vspace*{7.4cm}
 \hspace*{14.6cm}$\varrho$

\vspace*{-0.2cm}
 \hspace*{12.4cm}$\varrho_{ir}$

\vspace*{-0.5cm}
 \hspace*{5.2cm}$\varrho_{m}$

\vspace*{-5.64cm}
 \hspace*{7.7cm}$|\dot{\varphi}_{I}|$

\vspace*{3.4cm}
 \hspace*{13cm}$|\delta\dot{\varphi}_{\varrho}|\sim\frac{1}{\varrho^{2}}$

\vspace*{0.8cm}
 \hspace*{3.44cm}$\varrho_{uv}$

\vspace*{-0.5cm}
 \hspace*{2.8cm}$\varrho_{b}$

\vspace*{-7.28cm}
 \hspace*{0.9cm}$-\dot{\varphi}_{m}$

\vspace*{3.9cm}
 \hspace*{0.85cm}$-\dot{\varphi}_{uv}$

\vspace{1.5cm}

\vspace*{1cm}
\caption{Unlike the Coleman instantons, the new instanton, shown as a solid
black line, corresponds to the classical solution with the boundary
condition $\dot{\varphi}\left(\varrho_{b}\right)=0$ with finite non-vanishing
$\varrho_{b}$, which is a free parameter characterizing the entire
spectrum of the new instantons. Being continued to $\varrho<\varrho_{b}$,
this solution would become singular at $\varrho=0$ with $\varphi$
going to minus infinity. But already at $\varrho_{uv}>\varrho_{b}$
the derivative of the classical solution $|\dot{\varphi}_I|$ falls below the level 
of quantum fluctuations 
$|\delta\dot{\varphi}_{\varrho}|\sim\frac{1}{\varrho^{2}}$,
and $\varrho_{uv}$, which is determined by the parameters characterising
the classical solution, regularizes the singularity. For this reason,
the part of the solution indicated by the dotted red line in Fig.\ref{Figure2} 
must be ignored.}
\label{Figure4} 
\end{figure}

As can be seen in Fig.\ref{Figure4}, part of these new solutions inevitably
falls below the curve determining the level of quantum fluctuations,
and therefore they can be trusted only in the range $\varrho_{uv}<\varrho<\varrho_{ir}$
where $\dot{\varphi}\left(\varrho\right)$ exceeds the level of quantum
fluctuations. The ultraviolet cutoff scale $\varrho_{uv}$ is always
larger than $\varrho_{b}$. We will show how to determine this scale
for a rather broad class of potentials. To do this, we assume that
the field $\varphi$ does not change its value too much and stays
close to $\varphi_{b}$ when $\varrho$ changes from $\varrho_{b}$
to $\varrho_{uv}$ \footnote{For each concrete potential and solution this assumption must be verified
a posteriori and, as we have found, it is indeed valid in many cases. }. Then equation (\ref{5a-1}) can be well approximated as in (\ref{7a})
by replacing $\varphi_{c}$ by $\varphi_{b}$. The general solution
of this equation is given in (\ref{8a}), where the constant of integration
$C$ is now determined by the boundary condition (\ref{18}). As a
result, we find 
\begin{equation}
\varphi_{I}\left(\varrho\right)=\varphi_{b}+\frac{V_{b}'}{8}\,\frac{\left(\varrho^{2}-\varrho_{b}^{2}\right)^{2}}{\varrho^{2}}\,.\label{19}
\end{equation}



Substituting this solution into (\ref{12}), we obtain the following
equation: 
\begin{equation}
\varrho_{uv}^{4}+\frac{4\,\sigma}{V_{b}'}\,\varrho_{uv}-\varrho_{b}^{4}=0\,,\label{20}
\end{equation}
which can be solved exactly and gives us the ultraviolet cutoff scale
in terms of $\varrho_{b}$. The equation for the infrared cutoff
scale $\varrho_{ir}$ can be obtained in a similar way by considering
the asymptotics of the solution at large $\varrho$. At this stage,
we do not need explicit solutions for these scales but we would
like to emphasize again that the bounce for a new instanton always
occurs within the quantum core with the radius $\varrho_{uv}$. Quantum
fluctuations dominate in this core and it makes no sense to speak
about the classical solution for $\varrho<\varrho_{uv}$. Since we have
normalized the energy density in the false vacuum to zero, the energy
density of vacuum fluctuations in the quantum core must be shifted
by $V_{uv}\equiv V(\varphi_{uv})$ ($\varphi_{uv}\equiv 
\varphi(\varrho_{uv})$).
Taking this into account, we find the total potential energy
for $\tau=0$ which comprises both the contribution
of the classical instanton solution in its trustable region $\varrho_{uv}<\varrho<\varrho_{ir}$
and the energy of the quantum core: 
\begin{equation}
\mathcal{V}\left(0\right)=\frac{2\,\pi}{3}\left(\dot{\varphi}_{uv}^{2}\,\varrho_{uv}^{3}-\dot{\varphi}_{ir}^{2}\,\varrho_{ir}^{3}\right)\simeq\frac{2\,\pi\,\sigma}{3}\left(\frac{1}{\varrho_{uv}}-\frac{1}{\varrho_{ir}}\right).\label{21}
\end{equation}
Thus, with the precision allowed by the time-energy uncertainty relation,
the potential energy vanishes and the bubble with the quantum core emerges
from under the barrier, materialises and expands in the Minkowski space,
filling it with a new phase. Therefore, the after-bouncing part of
the new instanton at $0<\varrho<\varrho_{b}$, which is singular at
$\varrho\rightarrow0$, must be ignored and replaced by the quantum
core.

To determine the contribution of the new regularized instantons to
the decay rate, we have to calculate the Euclidean action (\ref{4a}),
which for the $O(4)$ solutions is simplified to
\begin{equation}
S_{E}\,=\,2\pi^{2}\,\int_{\varrho_{uv}}^{\varrho_{ir}}\left(\frac{1}{2}\,\dot{\varphi}^{2}\,+V(\varphi)\right)\varrho^{3}d\varrho+S_{QC}\,,\label{25}
\end{equation}
where 
\begin{equation}
S_{QC}=\frac{\pi^{2}}{2}V_{uv}\varrho_{uv}^{4}\,\label{26}
\end{equation}
is the contribution from the quantum core. Using equation of motion (\ref{5a-1})
and integrating by parts, we find 
\begin{align}
 & 2\pi^{2}\int_{\varrho_{uv}}^{\varrho_{ir}}V\varrho^{3}d\varrho=\frac{\pi^{2}}{2}\left(\left.V\varrho^{4}\right|_{\varrho_{uv}}^{\varrho_{ir}}-\int_{\varrho_{uv}}^{\varrho_{ir}}V'\dot{\varphi}\varrho^{4}d\varrho\right)\nonumber \\
 & =\frac{\pi^{2}}{2}\left(\left.V\varrho^{4}\right|_{\varrho_{uv}}^{\varrho_{ir}}-\int_{\varrho_{uv}}^{\varrho_{ir}}\left(\ddot{\varphi}+\frac{3}{\varrho}\dot{\varphi}\right)\dot{\varphi}\varrho^{4}d\varrho\right)\nonumber \\
 & =\frac{\pi^{2}}{2}\left(\left.\left(V-\frac{1}{2}\dot{\varphi}^{2}\right)\varrho^{4}\right|_{\varrho_{uv}}^{\varrho_{ir}}-\int_{\varrho_{uv}}^{\varrho_{ir}}\dot{\varphi}^{2}\varrho^{3}d\varrho\right)\,,\label{27}
\end{align}
and therefore total action (\ref{25}) can be rewritten as 
\begin{equation}
S_{E}=\frac{\pi^{2}}{2}\int_{\varrho_{uv}}^{\varrho_{ir}}\dot{\varphi}^{2}\varrho^{3}d\varrho+\frac{\pi^{2}}{4}\left(2V_{ir}\varrho_{ir}^{4}+\dot{\varphi}_{uv}^{2}\varrho_{uv}^{4}-\dot{\varphi}_{ir}^{2}\varrho_{ir}^{4}\right)\,.\label{28}
\end{equation}
The last term in the parentheses is of the order of unity and can be neglected
because the semiclassical approximation is valid only when $S_{E}\gg1$.
Moreover, 
\begin{equation}
\int_{\varrho_{b}}^{\varrho_{uv}}\dot{\varphi}^{2}\varrho^{3}d\varrho<\dot{\varphi}_{uv}^{2}\,\varrho_{uv}^{4}\simeq O(1)\,,\label{29}
\end{equation}
and the integral from $\varrho_{ir}$ to $\infty$ is also of the order of
unity. Therefore, the action for the new quantum-core instantons is
well approximated by 
\begin{equation}
S_{E}=\frac{\pi^{2}}{2}\int_{\varrho_{b}}^{\infty}\dot{\varphi}^{2}\varrho^{3}d\varrho\,.\label{30}
\end{equation}
The decay rate of the false vacuum \textit{per unit time per unit
volume} can then be estimated as 
\begin{equation}
\Gamma\simeq\varrho_{0}^{-4}\exp\left(-S_{E}\right)\,,\label{31}
\end{equation}
where $\varrho_{0}\equiv\varrho_{0}\left(\varrho_{b},\varphi_{f},...\right)$
is the parameter characterizing the ``size of the bubble'' and should
enter here for dimensionality reasons.

\section{General Solutions}

\begin{figure}[hbt]
\begin{centering}
\includegraphics[height=80mm]{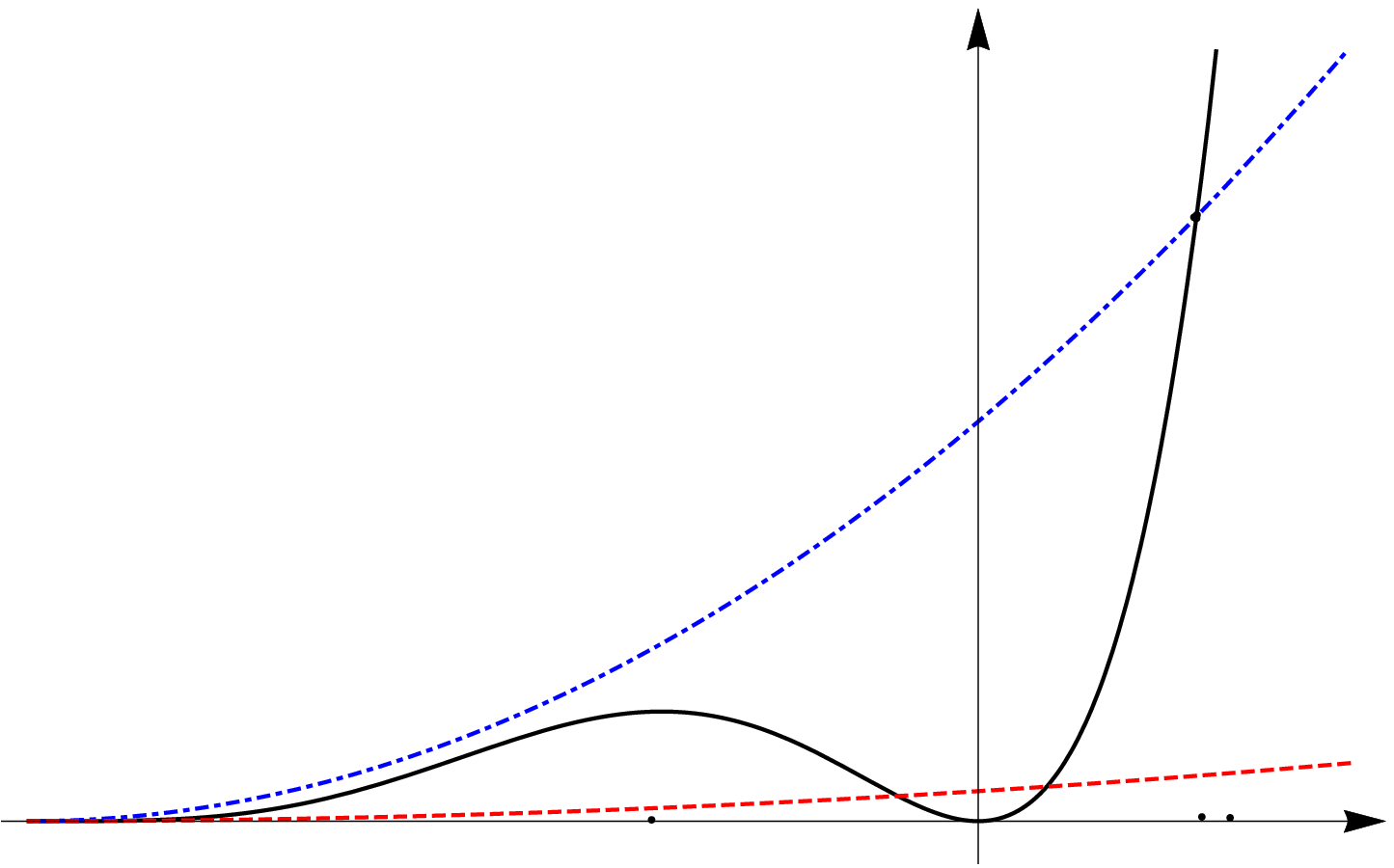} 
\par\end{centering}
\vspace*{-8.7cm}
 \hspace*{10.1cm} $|V'|$

\vspace*{7.39cm}
 \hspace*{14.55cm}$\varphi$

\vspace*{-0.3cm}
 \hspace*{13cm}$\varphi_{uv}$

\vspace*{-0.46cm}
 \hspace*{12.45cm}$\varphi_{m}$

\vspace*{-0.4cm}
 \hspace*{1.8cm}$\varphi_{f}$

\vspace*{-6.4cm}
 \hspace*{8.1cm} end of friction domination

\vspace*{4.4cm}
 \hspace*{11.8cm} thin-wall approximation
 
 \vspace*{0.5cm}
 \hspace*{7.4cm} $\varphi_{i}$

\vspace*{0.2cm}
\caption{The magnitude of the derivative of the potential is compared to the
value of the friction term in equation (\ref{5a-1}) calculated under
the assumption that this term dominates. In the case shown by the
blue dot-dashed line, the friction dominates over the potential
term in the range $\varphi_{f}<\varphi<\varphi_{m}$, where the derivative
$\left|\dot{\varphi}\right|$ grows as $\varrho$ changes from $\infty$
to $\varrho_{m}$ (see Fig.\ref{Figure4}). At $\varrho\simeq\varrho_{m}$ the
potential term in (\ref{5a-1}) takes over and $\left|\dot{\varphi}\right|$
starts to decrease until finally the ultraviolet cutoff is reached.
Such instantons are mostly dominated by friction and correspond to
thick-wall bubbles. If, on the other hand, the friction term is almost
completely below $\left|V'\right|$, as indicated by the dashed red
line, it can be neglected in the leading order and we can use the
thin-wall approximation.}
\label{Figure5} 
\end{figure}

In this section, we will find approximate instanton solutions for a
broad class of unbounded potentials with a false vacuum, as shown
in Fig.\ref{Figure1} . The false vacuum at $\varphi_{f}<0$ is separated from the
unbounded concave part of the potential by the barrier of hight $V_{bar}$,
and at $\varphi>0$ both the first and second derivatives of the potential
are negative, i.e. $V'<0$ and $V''<0$. The typical behavior of the
magnitude of the first derivative $\left|V'\left(\varphi\right)\right|$
for the potential in Fig.\ref{Figure1} is shown in Fig.\ref{Figure5}. It first grows, reaches
its maximum $V'_{i}$ at the inflection point located within the range
$\varphi_{f}<\varphi_{i}<0$, vanishes at $\varphi=0$ and then starts
to grow rapidly. We assume that the potential up to $\varphi_{i}$
can be well approximated as 
\begin{equation}
V\left(\varphi\right)=\frac{1}{2}V_{f}''\left(\varphi-\varphi_{f}\right)^{2}+\frac{1}{4!}V_{f}^{(IV)}\left(\varphi-\varphi_{f}\right)^{4}+...\,,\label{32}
\end{equation}
where $V''_{f}$ and $V_{f}^{(IV)}$ are the second and fourth derivatives
of the potential at $\varphi_{f}$, in particular, for the quartic potential
$V''_{f}=0$. Our task is to find an approximate solution of
equation (\ref{5a-1}) for a rather generic potential, simplifying
this equation by neglecting either the friction or the potential term
in it.

\vspace{0.5cm}


\subsection{Friction dominated solutions (outside of the thin-wall approximation)}

Given a solution of equation (\ref{5a-1}), we want to determine
in what range of $\varphi$ the potential term in this equation can
be neglected with respect to the friction term, i.e., 
\begin{equation}
\left|V'\right|\ll\left|\frac{3}{\varrho}\dot{\varphi}\right|\,.\label{33}
\end{equation}
To do this, we first find the following solution of equation (\ref{5a-1})
by omitting the potential term in it and taking into account the boundary
condition (\ref{6a}): 
\begin{equation}
\varphi\left(\varrho\right)\simeq\varphi_{f}-\frac{E}{4\,\varphi_{f}\,\varrho^{2}}\,,\label{34}
\end{equation}
where $E>0$ is the integration constant, normalized to negative $4\varphi_{f}$
for convenience because as we will see, $E$ is nothing more than the number
of quanta on the scale characterizing the size of the bubble. If
we calculate the derivative of $\varphi$ and express $\varrho$ in
terms of $\varphi$ from (\ref{34}), we get 
\begin{equation}
\left|\frac{3}{\varrho}\dot{\varphi}\right|\simeq\frac{24\left|\varphi_{f}\right|}{E}\left(\varphi_{f}-\varphi\right)^{2}\,.\label{35}
\end{equation}
It is clear that inequality (\ref{33}) is satisfied in the region
where $\left|V'\left(\varphi\right)\right|$ lies well below the quadraric
curve in (\ref{35}), (see Fig. 5). Therefore, for 
\begin{equation}
E\ll\frac{24\left|\varphi_{f}\right|}{V'_{i}}\left(\varphi_{f}-\varphi_{i}\right)^{2}\label{35a}
\end{equation}
the friction term dominates throughout the range from $\varphi_{f}$
to a \textit{positive} $\varphi_{m}$, determined by the equation
\begin{equation}
\frac{24\left|\varphi_{f}\right|}{E}\left(\varphi_{f}-\varphi_{m}\right)^{2}\simeq\alpha\left|V'_{m}\right|\,,\label{37}
\end{equation}
where $\alpha$ is a numerical coefficient of the order of unity and $V'_{m}\equiv V'\left(\varphi_{m}\right)$.
As can be seen from Fig.\ref{Figure5}, this equation always has a solution for
any $E$ when $\left|V'\right|$ grows faster than $\varphi^{2}$
for positive $\varphi$ \footnote{Note that if the potential is quadratic near a false vacuum, there
is a small region around $\varphi_{f}$ of size $\Delta\varphi/\left|\varphi_{f}\right|\simeq E\,V''_{f}/\varphi_{f}^{2}\ll1$
where the potential term is important. However, if the condition (\ref{35a})
is satisfied, the contribution of this region is negligible.}. In many cases $V'_{i}$ can be estimated as $O(1)\times V_{bar}/\left|\varphi_{f}\right|$
and $\left|\varphi_{f}-\varphi_{i}\right|\sim\left|\varphi_{f}\right|$
and inequality (\ref{35a}) is simplified to 
\begin{equation}
E\ll\frac{O(1)\,\varphi_{f}^{4}}{V_{bar}}\,.\label{36-1}
\end{equation}
When inequality (\ref{35a}) or its simplified version (\ref{36-1})
is satisfied, the instanton solution in the range $\varphi_{f}<\varphi<\varphi_{m}$
is well approximated by (\ref{34}). The field $\varphi$ changes
from negative to positive values as $\varrho$ decreases, and it
vanishes at 
\begin{equation}
\varrho_{0}^{2}\simeq\frac{E}{4\,\varphi_{f}^{2}}\,\label{38}
\end{equation}
characterizing the ``size of the bubble''. The number of quanta
at a scale of $\varrho_{0}$ is of the order 
\begin{equation}
\dot{\varphi}_{0}^{2}\,\varrho_{0}^{4}\simeq\frac{E^{2}}{4\,\varphi_{f}^{2}\,\varrho_{0}^{2}}\simeq E\,,\label{39}
\end{equation}
and this number of quanta $E$ must be much larger than unity, because
otherwise the essential part of the classical instanton is completely
saturated by quantum fluctuations. As follows from (\ref{38}), the
size of the instanton $\varrho_{0}$ is always larger than $1/\left|\varphi_{f}\right|$
and the magnitude of the negative ``velocity'' at the $\varphi=0$ crossing, 
\begin{equation}
\dot{\varphi}(\varrho_{0})\simeq-\frac{4\,\varphi_{f}^{2}}{E^{1/2}}\,,\label{40}
\end{equation}
is smaller than $\varphi_{f}^{2}$. After the field becomes positive
and $\varrho$ continues to decrease, the friction term dominates
until $\varphi$ approaches $\varphi_{m}$ at 
\begin{equation}
\varrho_{m}^{2}\simeq\frac{E}{4\,\varphi_{f}\left(\varphi_{f}-\varphi_{m}\right)}\,,\label{41}
\end{equation}
where $\varphi_{m}$ is the solution of equation (\ref{37}). At this
``moment'', the potential term in equation (\ref{5a-1}) becomes
comparable to the friction term, the magnitude of the field derivative
$\left|\dot{\varphi}\right|$ reaches its maximum 
\begin{equation}
\left|\dot{\varphi}_{m}\right|\simeq\sqrt{\frac{E\left(\varphi_{f}-\varphi_{m}\right)}{\varphi_{f}}}\varrho_{m}^{-2}\gg\varrho_{m}^{-2}\,,\label{41-2-1}
\end{equation}
and for $\varrho<\varrho_{m}$, where the potential term dominates,
$\left|\dot{\varphi}\right|$ decreases vanishing at the bounce at
$\varrho_{b}<\varrho_{m}$. As can be seen in Fig.\ref{Figure4}, just before the
bounce $\left|\dot{\varphi}\left(\varrho_{uv}\right)\right|$ becomes
comparable to the level of quantum fluctuations, and the bubble with
a quantum core emerges in the Minkowski space. To estimate the potential
in the quantum core $V_{uv}\equiv V(\varphi(\varrho_{uv}))$, we
omit the friction term in equation (\ref{5a-1}) at $\varrho<\varrho_{m}$
and take into account that $\dot{\varphi}_{uv}^{2}\ll\dot{\varphi}_{m}^{2}$.
Then from the first integral of this simplified equation we find 
\begin{equation}
V_{uv}\simeq V_{m}-\frac{1}{2}\dot{\varphi}_{m}^{2}\simeq V_{m}-\frac{8\,\varphi_{f}\left(\varphi_{f}-\varphi_{m}\right)^{3}}{E}\,,\label{42}
\end{equation}
where $V_{m}\equiv V(\varphi_{m})$. Taking into account
(\ref{36-1}), we find by considering this expression that the maximal
value of the potential in the quantum core for the friction dominated
instantons is about $-V_{bar}$ and the quantum bound $E>1$ imposes
a lower bound on the depth of penetration under the barrier. 

Now let
us estimate the action for the friction dominated instantons. Taking
into account that $\dot{\varphi}^{2}$ decreases for $\varrho<\varrho_{m}$,
we find that 
\begin{equation}
\int_{\varrho_{b}}^{\varrho_{m}}\dot{\varphi}^{2}\varrho^{3}d\varrho\apprle\frac{1}{4}\dot{\varphi}_{m}^{2}\varrho_{m}^{4}\simeq\frac{E\,\left(\varphi_{f}-\varphi_{m}\right)}{4\,\varphi_{f}}\,.\label{43}
\end{equation}
In the range $\varrho_{m}<\varrho<\infty$ the friction term dominates
and the solution is well approximated by (\ref{34}), thus 
\begin{equation}
\int_{\varrho_{m}}^{\infty}\dot{\varphi}^{2}\varrho^{3}d\varrho\simeq\frac{E\,\left(\varphi_{f}-\varphi_{m}\right)}{2\,\varphi_{f}},\label{44}
\end{equation}
and the total action (\ref{30}) is about 
\begin{equation}
S_{E}\simeq\frac{3\,\pi^{2}}{8}\frac{E\,\left(\varphi_{f}-\varphi_{m}\right)}{\varphi_{f}}\,.\label{45}
\end{equation}

\subsection{The thin-wall approximation}

For 
\begin{equation}
E\gg\frac{O(1)\,\varphi_{f}^{4}}{V_{bar}}\,,\label{46}
\end{equation}
the magnitude of the negative potential in the quantum core $\left|V_{uv}\right|$
is much smaller than the hight of the potential $V_{bar}$. In this
case, we can use the thin-wall approximation, which assumes that the
instanton is dominated by the potential term in equation (\ref{5a-1})
and the friction term can be considered as a small perturbation. Let
us check whether this is indeed the case. The exact non-local first
integral of equation (\ref{5a-1}) is 
\begin{equation}
\frac{1}{2}\,\dot{\varphi}^{2}-V=\int_{\varrho}^{\infty}\frac{3}{\tilde{\varrho}}\,\dot{\varphi}^{2}\,d\tilde{\varrho},\label{47}
\end{equation}
where the boundary condition (\ref{6a}) has been used. We now assume
that the field $\varphi$ changes very rapidly only within the thin
layer of the width $2\Delta\varrho$, which is much smaller than the size
of the bubble $\varrho_{0}$, and accordingly tends to $\varphi_{f}$
and $\varphi_{uv}$ rapidly for $\varrho>\varrho_{0}$ and $\varrho<\varrho_{0}$, respectively.
Then the integral on the right-hand side of (\ref{47}) is suppressed
within the wall by a factor $\Delta\varrho/\varrho_{0}$ compared
to $\dot{\varphi}^{2}$ and $V$. Therefore, in the leading approximation
we have 
\begin{equation}
\dot{\varphi}^{2}\approx 2\,V\label{48}
\end{equation}
for $\varrho_{0}-\Delta\varrho<\varrho<\varrho_{0}+\Delta\varrho$.
Recalling that $\varphi\left(\varrho_{0}\right)=0$, we immediately
find that the size of the bubble can be estimated as 
\begin{equation}
\varrho_{0}\simeq\left(\frac{E}{2\,V_{bar}}\right)^{1/4}\,,\label{49}
\end{equation}
where $E\equiv\dot{\varphi}_{0}^{2}\,\varrho_{0}^{4}$ is the number
of quanta on the bubble scale. If we neglect $\dot{\varphi}_{uv}^{2}$
compared to $V_{uv}$, we find from (\ref{47}) 
\begin{equation}
V_{uv}\approx-\frac{3\,s}{\varrho_{0}}\,,\label{50}
\end{equation}
where 
\begin{equation}
s\equiv\int_{\varrho_{uv}}^{\infty}\dot{\varphi}^{2}d\varrho\approx\int_{\varphi_{f}}^{\varphi_{uv}}\sqrt{2\,V}d\varphi\sim\sqrt{2\,V_{bar}}\,\left(\varphi_{uv}-\varphi_{f}\right)\label{51}
\end{equation}
is the surface tension of the bubble. The value of $\varphi_{uv}>0$
is determined in the leading order by $V\left(\varphi_{uv}\right)\approx0.$
From (\ref{49}), (\ref{50}) and (\ref{51}) we obtain 
\begin{equation}
V_{uv}\simeq-\frac{3\left(2\,V_{bar}\right)^{3/4}\left(\varphi_{uv}-\varphi_{f}\right)}{E^{1/4}},\label{52}
\end{equation}
and it is obvious that for those $E$ satisfying inequality (\ref{46})
we have $\left|V_{uv}\right|\ll V_{bar}$. The relative width of the
bubble wall can be evaluated as $\Delta\varrho/\varrho_{0}\sim\left|V_{uv}\right|/V_{bar}$.
Finally, the action (\ref{30}) can be approximated as 
\begin{equation}
S_{E}\simeq\frac{\pi^{2}}{2}\,\varrho_{0}^{3}\,s\simeq\frac{\pi^{2}}{2}\frac{E^{3/4}}{\left(2\,V_{bar}\right)^{1/4}}\left(\varphi_{uv}-\varphi_{f}\right)\,.\label{53}
\end{equation}
We would like to emphasize again that the numerical coefficients in
all the above formulas are determined up to a factor of the order of unity.

\subsection{Summary} 

For convenience, let us briefly summarize the basic
formulas obtained in this Section. As we have found, quantum fluctuations
regularize the classical solutions which would otherwise be singular.
This in turn leads to the appearance of a whole class of new instantons,
all of which contribute to the false vacuum decay. These new instantons
can be parametrized by the number of quanta $E\equiv\dot{\varphi}_{0}^{2}\varrho_{0}^{4}$
on the scale of the instanton size $\varrho_{0}$. As a function of
$E$, we derived the following approximate formulas characterizing
the instantons for a general unbounded potential in Fig.1, which is
concave at positive $\varphi$.

For the {\it friction dominated instantons}, for which
\begin{equation}
1\ll E\ll\frac{O\left(1\right)\varphi_{f}^{4}}{V_{bar}}\,,\label{54}
\end{equation}
the bubble size, the potential in the quantum core, and the instanton action are given
accordingly 
\begin{equation}
\varrho_{0}\simeq\frac{E^{1/2}}{2\left|\varphi_{f}\right|}\,,\quad V_{uv}\simeq V\left(\varphi_{m}\right)-\frac{8\,\varphi_{f}\left(\varphi_{f}-\varphi_{m}\right)^{3}}{E}\,,\quad S_{E}\simeq\frac{3\,\pi^{2}}{8}\frac{E\left(\varphi_{f}-\varphi_{m}\right)}{\varphi_{f}}\,,\label{55}
\end{equation}
where $\varphi_{f}<0$ is the location of the false vacuum, $V_{bar}$
is the hight of the barrier, and $\varphi_{m}$ is the solution of
the equation 
\begin{equation}
\frac{24\,\left|\varphi_{f}\right|}{E}\left(\varphi_{f}-\varphi_{m}\right)^{2}\simeq O(1)\left|V'\left(\varphi_{m}\right)\right|\,.\label{56}
\end{equation}

In the case of {\it the thin-wall approximation}, for 
\begin{equation}
E\gg\frac{O\left(1\right)\varphi_{f}^{4}}{V_{bar}}\,,\label{57}
\end{equation}
we have 
\begin{equation}
\varrho_{0}\simeq\left(\frac{E}{2\,V_{bar}}\right)^{1/4}\,,\quad V_{uv}\simeq-\frac{3\left(2\,V_{bar}\right)^{3/4}\left(\varphi_{uv}-\varphi_{f}\right)}{E^{1/4}}\,,\quad S_{E}\simeq\frac{\pi^{2}}{2}\frac{E^{3/4}}{\left(2\,V_{bar}\right)^{1/4}}\left(\varphi_{uv}-\varphi_{f}\right)\,,\label{58}
\end{equation}
where $\varphi_{uv}$ is the solution of the equation $V\left(\varphi_{uv}\right)\approx0.$

Note that these formulas also apply well for the potential with the
second true minimum $V_{true}<0$ at a positive $\varphi_{tr}$, if
the potential up to this minimum is well approximated by a concave
potential, as shown, e.g., by the dot-dashed line in Fig.\ref{Figure1}. In this case,
the minimal value of $E$ is determined by $V_{uv}(E_{min})\simeq V_{true}$
if $E_{min}>1$, otherwise it must be assumed to be of the order of unity if
the true minimum is too low. Given $\varrho_{0}(E)$ and $S_{E}\left(E\right)$,
the contribution of the corresponding instanton parametrized by $E$
to the decay rate per unit time per unit volume can be estimated as
$\Gamma\simeq\varrho_{0}^{-4}\exp\left(-S_{E}\right)$.

\section{Examples}

We will now show how the above results can be applied to the general
class of potentials and derive concrete formulas in several interesting
examples. We will also compare these results with those obtained from
exact solutions in the cases where such solutions exist.

\subsection{Concave power-law potentials} 

Let us start with the class
of unbounded potentials for which the Coleman instantons do not
exist \cite{MS}. Namely, let us consider the potential 
\begin{equation}
V\left(\varphi\right)=\left\{ \begin{array}{cc}
\frac{\lambda_{+}}{4}\left(\varphi-\varphi_{f}\right)^{4} & \text{for }\varphi<\beta\varphi_{f}\\
-\frac{\lambda_{-}}{n}\varphi_{f}^{4}\left(\frac{\varphi}{\varphi_{f}}\right)^{n}+V_{bar} & \text{for }\varphi>\beta\varphi_{f}
\end{array}\right.\label{59}
\end{equation}
with $n\geq4$, where the parameter $\beta$ is expressed in
terms of the positive coupling constants $\lambda_{+}$ and $\lambda_{-}$ as
\begin{equation}
\frac{\beta^{n-1}}{\left(1-\beta\right)^{3}}=\frac{\lambda_{+}}{\lambda_{-}}\label{60}
\end{equation}
and 
\begin{equation}
V_{bar}=\left[\frac{\lambda_{-}}{n}\beta^{n}+\frac{\lambda_{+}}{4}\left(1-\beta\right)^{4}\right]\varphi_{f}^{4}\label{61}
\end{equation}
is the hight of the potential. This potential is composed of two power-law
potentials which meet at $\varphi=\beta\varphi_{f}$, and the first
derivative of the potential is continuous at this point. We consider
two limiting cases, namely, when the potential drops very rapidly
after reaching its maximum ($\lambda_{-}\gg\lambda_{+}$) and when
the potential is very flat near the maximum ($\lambda_{-}\ll\lambda_{+}).$

\textbf{\textit{a)}}{ \textit{\bf \textit{The potential with a sharp maximum}}\textbf{
}($\lambda_{-}\gg\lambda_{+}$). In this case, $\beta\simeq(\lambda_{+}/\lambda_{-})^{1/\left(n-1\right)}\ll1$
and 
\begin{equation}
V_{bar}\simeq\frac{\lambda_{+}}{4}\varphi_{f}^{4}\,.\label{61-1}
\end{equation}
We first consider the friction-dominated instantons with 
\begin{equation}
1\ll E\ll\frac{1}{\lambda_{+}}\,.\label{62}
\end{equation}
Solving equation (\ref{56}) for $\varphi_{m}$, we find that for
$1\ll E\ll1/\lambda_{-}$ 
\begin{equation}
\varphi_{m}\sim\frac{\varphi_{f}}{\left(\lambda_{-}E\right)^{1/\left(n-3\right)}},\;\;V_{uv}\sim-\frac{\lambda_{-}\varphi_{f}^{4}}{\left(\lambda_{-}E\right)^{n/\left(n-3\right)}},\;\;S_{E}\sim\frac{E}{\left(\lambda_{-}E\right)^{1/\left(n-3\right)}},\label{63a}
\end{equation}
while for $1/\lambda_{-}\ll E\ll1/\lambda_{+}$ one obtains 
\begin{equation}
\varphi_{m}\sim\frac{\varphi_{f}}{\left(\lambda_{-}E\right)^{1/\left(n-1\right)}}\,,\quad V_{uv}\sim-\frac{\varphi_{f}^{4}}{E}\,,\quad S_{E}\sim E\,.\label{64a}
\end{equation}
The size of the bubble in both cases is 
\begin{equation}
\varrho_{0}\sim\frac{E^{1/2}}{\left|\varphi_{f}\right|}\,.\label{66}
\end{equation}
In the above formulas, we have completely omitted the numerical coefficients
leaving only the parametric dependence.

In the thin-wall approximation for 
\begin{equation}
E\gg\frac{1}{\lambda_{+}}\,\label{67}
\end{equation}
from (\ref{58}) we immediately find that for any $n\geq4$ 
\begin{equation}
\varrho_{0}\sim\frac{1}{\left|\varphi_{f}\right|}\left(\frac{E}{\lambda_{+}}\right)^{1/4}\,,\quad V_{uv}\sim-\frac{V_{bar}}{\left(\lambda_{+}E\right)^{1/4}}\,,\quad S_{E}\sim\frac{\left(\lambda_{+}E\right)^{3/4}}{\lambda_{+}}\,.\label{68}
\end{equation}

\textbf{\textit{b)}}{ \textit{\bf \textit{The potential with a flat maximum}}\textbf{
}
In the case $\lambda_{-}\ll\lambda_{+}$ the hight of the barrier
is 
\begin{equation}
V_{bar}\simeq\frac{\lambda_{-}}{n}\varphi_{f}^{4}\,.\label{69}
\end{equation}

For the friction-dominated instantons with $1\ll E\ll\frac{1}{\lambda_{-}}$,
the solution of equation (\ref{56}) is 
\begin{equation}
\varphi_{m}\sim\frac{\varphi_{f}}{\left(\lambda_{-}E\right)^{1/\left(n-3\right)}}\gg\varphi_{f}
\label{70}
\end{equation}
and formulas (\ref{55}) are simplified to 
\begin{equation}
\varrho_{0}\sim\frac{E^{1/2}}{\left|\varphi_{f}\right|}\,,\quad V_{uv}\sim-\frac{V_{bar}}{\left(\lambda_{-}E\right)^{n/\left(n-3\right)}}\,,\quad S_{E}\sim\frac{E}{\left(\lambda_{-}E\right)^{1/\left(n-3\right)}}\,.\label{71}
\end{equation}

For $E\gg\frac{1}{\lambda_{-}}$ (the thin-wall approximation) it follows from
(\ref{58}) that 
\begin{equation}
\varrho_{0}\sim\frac{1}{\left|\varphi_{f}\right|}\left(\frac{E}{\lambda_{-}}\right)^{1/4}\,,\quad V_{uv}\sim-\frac{V_{bar}}{\left(\lambda_{-}E\right)^{1/4}}\,,\quad S_{E}\sim\frac{\left(\lambda_{-}E\right)^{3/4}}{\lambda_{-}}\,.\label{72}
\end{equation}

In the case of the quartic unbounded potential $\left(n=4\right)$,
the exact solution was constructed in terms of the Jacobi elliptic functions
and the corresponding asymptotic expressions were derived in \cite{MRS2}.
The reader can verify that the above asymptotic formulas are in full
agreement with the results of \cite{MRS2}. Thus, we have shown that
the approximate formulas derived in \cite{MRS2}, as a result of rather
lengthy calculations, can be obtained in a very simple way using the
``friction-dominated'' and ``thin-wall'' approximations developed
in this work\footnote{There is a mistake in the paper \cite{MRS2}, where the first term
in the action (45) there, which is proportional to $E_{-}$, must
be absent.}. Moreover, for $n>4$ the exact solutions for the instantons do not
exist, while with our approach we can find approximate solutions
describing the tunneling for any $n.$

\vspace{0.5cm}


\begin{figure}[hbt]
\begin{centering}
\includegraphics[height=80mm]{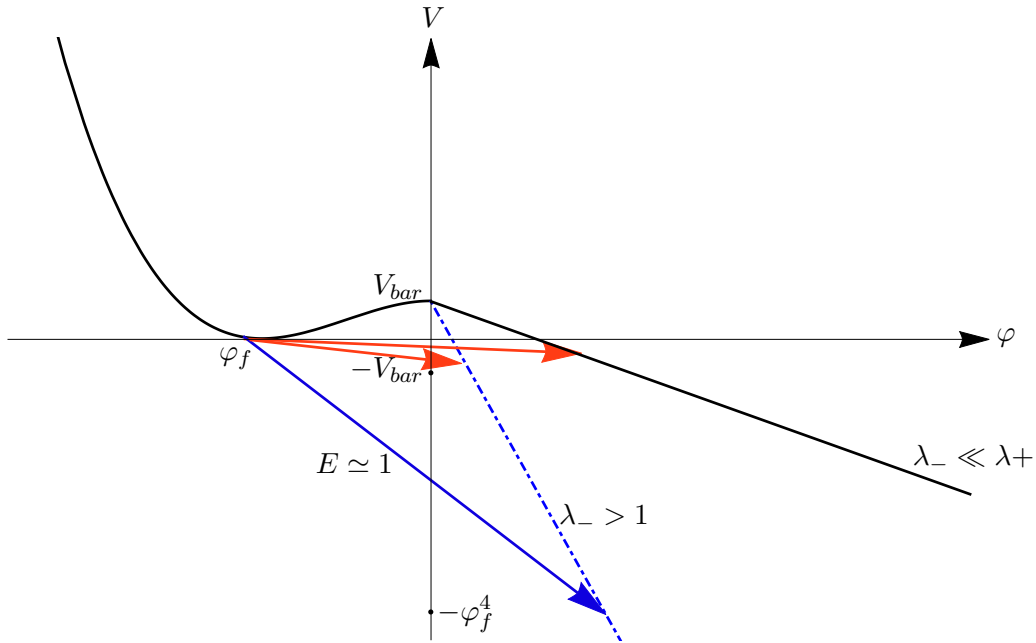} 
\par\end{centering}
\vspace*{-4.2cm}
 \hspace*{4.2cm} $\varphi_{f}$

\vspace*{-4.9cm}
 \hspace*{6.99cm}$V$

\vspace*{3.69cm}
 \hspace*{14.55cm}$\varphi$

\vspace*{0cm}
 \hspace*{6.05cm}$-V_{bar}$

\vspace*{0.8cm}
 \hspace*{5.6cm}$E\simeq1$

\vspace*{1.46cm}
 \hspace*{7.2cm}$-\varphi_{f}^{4}$

\vspace*{-4.8cm}
 \hspace*{6.35cm}$V_{bar}$

\vspace*{1.8cm}
 \hspace*{13.45cm}$\lambda_{-}\ll\lambda+$

\vspace*{0.3cm}
 \hspace*{8.8cm}$\lambda_{-}>1$

\vspace{1.5cm}

\vspace*{0.0cm}
\caption{In the case of the flat linear potential with $\lambda_{-}\ll\lambda_{+}$
only the thin-wall instantons with $\left|V_{uv}\right|<\left|V_{bar}\right|$
represented by the red solid line exist. The limiting case $\left|V_{uv}\right|\simeq\left|V_{bar}\right|$
corresponds to the Coleman instanton. For the steep potential with
$\lambda_{-}>1$, both the friction-dominated and thin-wall instantons
exist. The lower blue line shows the transition for the friction-dominated
instanton to the lowest possible value of the potential in the quantum
core corresponding to $E\simeq1$. For the Coleman instanton, which
also exists in this case, $E\ll1$, and thus it is completely saturated
by quantum fluctuations.}
\label{Figure6} 
\end{figure}

\vspace{0cm}

\subsection{Linear potential} 

The methods developed in this work can be applied with an obvious
modification to a much broader class of potentials, namely, the
potentials with two minima and the potentials which are not necessarily
concave. To demonstrate how this can be done in cases other than those
presented above, now we consider the linear potential
\begin{equation}
V\left(\varphi\right)=\left\{ \begin{array}{cc}
\frac{\lambda_{+}}{4}\left(\varphi-\varphi_{f}\right)^{4} & \text{for }\varphi<0\\
\lambda_{-}\varphi_{f}^{3}\varphi+V_{bar} & \text{for }\varphi>0\,,
\end{array}\right.\label{73}
\end{equation}
for which we can compare the approximate expressions derived by the methods of this work with the
results obtained from exact solutions constructed in \cite{MRS1}.
Here, $\varphi_{f}<0$, the coupling constants  $\lambda_{+}$ and $\lambda_{-}$ are positive, the potential has the hight 
\begin{equation}
V_{bar}=\frac{1}{4}\lambda_{+}\varphi_{f}^{4}\label{74}
\end{equation}
and the first derivative of the potential is discontinuous at $\varphi=0$ (see Fig.\ref{Figure6}). 
Depending on the ratio of the coupling constants, we can have
different types of instantons and therefore we consider two limiting
cases separately. As before, we will skip all numerical coefficients
and focus on the parametric dependence of the instanton solutions.

\textbf{\textit{a)}}{ \textit{\bf \textit{ Flat potential}}\textbf{
}($\lambda_{-}\ll\lambda_{+})$.
In this case, the magnitude of the derivative of the potential $\left|V'\right|$
as a function of $\varphi$ is shown in Fig.\ref{Figure7}. As can be seen from
this figure for $E<1/\lambda_{+}$, the friction term (see the green dotted
curve) dominates everywhere and the corresponding instantons cannot
be regularized by quantum fluctuations. In fact, for them the time
derivative $\left|\dot{\varphi}\right|$ behaves as $\varrho^{-3}$
at small $\varrho$ and the classical field dominates over quantum
fluctuations everywhere up to a singularity at $\varrho=0$. Therefore,
these instantons do not contribute to the vacuum decay rate. For the non-singular
solutions, the potential always dominates at negative $\varphi$ and
we can apply formulas (\ref{49}) to estimate $\varrho_{0}$. There
is a jump of the derivative at $\varphi=0$; it becomes much smaller
at $\varphi=+0$ as it was at $\varphi=-0$. If we want to avoid
the dominance of the friction term directly after the $\varphi=0$ crossing,
which would lead us to the singularity, we must require that 
\begin{equation}
\frac{3}{\varrho_{0}}\left|\dot{\varphi_{0}}\right|\sim\frac{V_{bar}^{3/4}}{E^{1/4}}\leq\left|V_{+0}'\right|=\left|\lambda_{-}\varphi_{f}^{3}\right|\,,\label{75}
\end{equation}
where we have used equations (\ref{48}) and (\ref{49}). It follows from here that only for 
\begin{equation}
E\geq E_{C}\simeq\frac{\lambda_{+}^{3}}{\lambda_{-}^{4}}\label{76}
\end{equation}
the instantons are non-singular. These instantons fulfil the thin-wall
criterion and therefore one can apply formulas (\ref{58}), which
in this case lead to
\begin{equation}
\varrho_{0}\sim\frac{1}{\left|\varphi_{f}\right|}\left(\frac{E}{\lambda_{+}}\right)^{1/4},\quad V_{uv}\sim-V_{bar}\left(\frac{E_{C}}{E}\right)^{1/4},\quad S_{E}\sim E_{C}\left(\frac{E}{E_{C}}\right)^{3/4},\label{77}
\end{equation}
where we have taken into account that $\varphi_{uv}\simeq\left(\lambda_{+}/\lambda_{-}\right)\varphi_{f}$
as it follows from $V\left(\varphi_{uv}\right)\simeq0.$ The minimum
value $E=E_{C}$ corresponds to the instanton with the Coleman boundary
condition that exists in this case.

\vspace{0.5cm}


\begin{figure}[hbt]
\begin{centering}
\includegraphics[height=80mm]{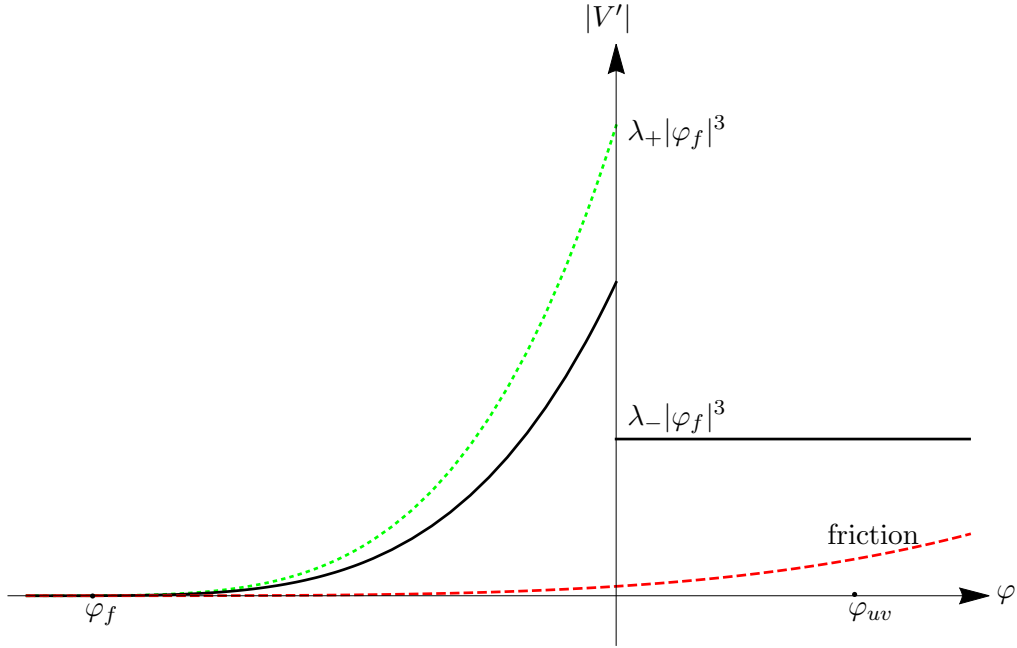} 
\par\end{centering}

\vspace*{-8.7cm}
 \hspace*{9.0cm} $|V'|$

\vspace*{7.1cm}
 \hspace*{14.55cm}$\varphi$

\vspace*{-0.25cm}
 \hspace*{12.6cm}$\varphi_{uv}$

\vspace*{-0.46cm}
 \hspace*{2.56cm}$\varphi_{f}$

\vspace*{-6.8cm}
 \hspace*{9.7cm}$\lambda_{+}|\varphi_{f}|^{3}$

\vspace*{3.3cm}
 \hspace*{9.7cm}$\lambda_{-}|\varphi_{f}|^{3}$

\vspace*{1.1cm}
 \hspace*{12.2cm} friction

\vspace*{1.0cm}
\caption{The green dotted line describes the singular instantons for which
the friction term dominates for both negative and positive values of $\varphi$.
For non-singular instantons the friction term calculated in (\ref{35})
under the assumption that it dominates (see the red dashed line) is always
below the $\left|V'\right|$-curve at negative $\varphi$ and it remains
less relevant than the potential at positive $\varphi$ only
if the condition (\ref{76}) is fulfiled.}
\label{Figure7} 
\end{figure}

\textbf{\textit{a)}}{ \textit{\bf \textit{Steep potential}}\textbf{
}
($\lambda_{+}\ll\lambda_{-})$.
The magnitude of the derivitave of the potential for this case is
shown in Fig.8. The derivative also jumps at $\varphi=0,$ but unlike
to the previous case it becomes much larger at $\varphi=+0$ than
it is at $\varphi=-0$. Therefore, here there exist nonsingular solutions
describing instantons dominated by friction at $\varphi<0.$ Taking
into account (\ref{35}), we can see from Fig.\ref{Figure8} that the instantons
with $E<1/\lambda_{-}$ are also dominated by friction at $\varphi>0$
and are therefore singular. The non-singular instantons with 
\begin{equation}
\frac{1}{\lambda_{-}}\ll E\ll\frac{1}{\lambda_{+}}\,,\label{78}
\end{equation}
are then described by expressions in (\ref{55}), where we have
to set $\varphi_{m}=0$: 
\begin{equation}
\varrho_{0}\sim\frac{E^{1/2}}{\left|\varphi_{f}\right|}\,,\quad V_{uv}\sim-\frac{\varphi_{f}^{4}}{E}\,,\quad S_{E}\sim E\,.\label{79}
\end{equation}
The minimal value $E\simeq1/\lambda_{-}$ corresponds to the Coleman
instanton. Note that for a very steep potential with $\lambda_{-}>1$
the Coleman instanton with $E<1$ is entirely saturated by quantum
fluctuations. In this case, in formulas (\ref{79}) $E$ must
be taken greater than unity and thus the minimal value of the potential
in the quantum core is about $V_{uv}\sim-\varphi_{f}^{4}$ and the
minimal size of the bubble, which should be the main contributor to
the decay rate for the unbounded potential, is $\varrho_{0}\sim1/\left|\varphi_{f}\right|$.



\begin{figure}[hbt]
\begin{centering}
\includegraphics[height=80mm]{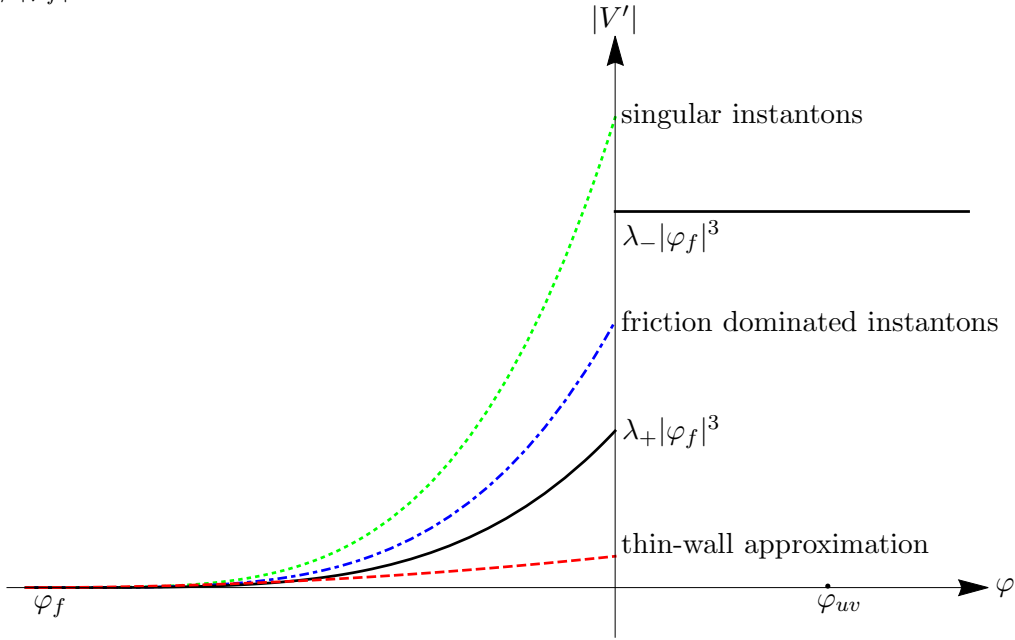} 
\par\end{centering}
\vspace*{-8.58cm}
 \hspace*{9.1cm} $|V'|$

\vspace*{7cm}
 \hspace*{14.55cm}$\varphi$

\vspace*{-0.27cm}
 \hspace*{12.2cm}$\varphi_{uv}$

\vspace*{-0.46cm}
 \hspace*{1.9cm}$\varphi_{f}$

\vspace*{-5.3cm}
 \hspace*{9.64cm}$\lambda_{-}|\varphi_{f}|^{3}$

\vspace*{2.1cm}
 \hspace*{9.64cm}$\lambda_{+}|\varphi_{f}|^{3}$

\vspace*{-4.67cm}
 \hspace*{9.5cm} singular instantons

\vspace*{2.29cm}
 \hspace*{9.5cm} friction dominated instantons

\vspace*{2.46cm}
 \hspace*{9.5cm} thin-wall approximation

\vspace*{0.9cm}
\caption{For very steep linear potentials with $\lambda_{-}>1$ we have, depending
on $E$, the singular, the friction-dominated and the thin-wall instantons.
They correspond to the dotted green, dot-dashed blue and dashed red curves,
respectively. The Coleman instanton, which also exists in this case,
does not belong to the spectrum because $E<1$ and this instanton
is thus completely saturated by quantum fluctuations.}
\label{Figure8} 
\end{figure}

For $E\gg1/\lambda_{+}$, the thin-wall formulas (\ref{58}) are valid
and taking into account that $\varphi_{uv}\simeq\left(\lambda_{+}/\lambda_{-}\right)\varphi_{f}\ll\varphi_{f}$,
we obtain 
\begin{equation}
\varrho_{0}\sim\frac{1}{\left|\varphi_{f}\right|}\left(\frac{E}{\lambda_{+}}\right)^{1/4}\,,\quad V_{uv}\sim-\frac{V_{bar}}{\left(\lambda_{+}E\right)^{1/4}}\,,\quad S_{E}\sim\frac{\left(\lambda_{+}E\right)^{3/4}}{\lambda_{+}}\,.\label{80}
\end{equation}
One can easily verify that the above formulas reproduce all the limiting
cases derived from the exact solutions in \cite{MRS1} \footnote{In order to compare the results obtained in this work with those in
\cite{MRS1}, one needs to use the following relation between the parameter
$E$ and the parameter $\chi$ introduced in \cite{MRS1}: 
$
E=\frac{32}{\lambda_{-}\,\left(1-\beta\right)^{3}}\left(1+\chi\right)\left(1+\beta\,\chi\right)^{3},
$
where $\beta\equiv\lambda_{+}/\left(\lambda_{+}+\lambda_{-}\right).$}.

\section{Conclusions}

In this work, we have shown that quantum fluctuations play an important
role when we consider the decay of a false vacuum. In particular,
they regularize the classical instanton solutions, which would otherwise
be singular. This in turn leads to the appearance of the full spectrum
of new instantons, all of which contribute to the decay of the false
vacuum. In those cases, when instantons with the Coleman boundary conditions
exist, they normally belong to this spectrum. However, as we
have seen in the example of the steep linear potential, there are
potentials where the Coleman instantons are completely saturated by
the quantum fluctuations and thus are no longer trustworthy as classical
solutions. Moreover, we proved in \cite{MS} that for a broad class
of potentials, where the false vacuum must obviously be unstable,
the Coleman instantons do not exist at all. For these potentials, the decay
of the vacuum is fully due to our new instantons,
which actually are bouncing solutions regularized by the ultraviolet
cutoff, which is self-consistently expressed in terms of the new instanton parameters.

The complete set of new instantons are classical solutions of equation
(\ref{5a-1}) with the bounce boundary conditions $\dot{\varphi}\left(\varrho_{b}\right)=0$
and $\varphi\left(\varrho\rightarrow\infty\right)=\varphi_{f}$, which are reliable only within the range $\varrho_{uv}<\varrho<\varrho_{ir}$,
where the ultraviolet and infrared cutoff scales are two different solutions of
equation (\ref{12}). At $\varrho<\varrho_{uv}$, the bubble is
filled by dominant vacuum fluctuations with the ground state level shifted
to $V_{uv}$. The proposed boundary conditions comprises a free parameter $\varrho_{b}$, 
which parametrizes the new instantons,
however they in fact were parametrized by the parameter $E$ instead of $\varrho_{b}$, 
which is a number of quanta on the bubble size scale, i.e. $E\equiv\dot{\varphi}^{2}\varrho_{0}^{4}$, 
and it is more suitable for the purpose of the present paper. It is clear
that for any given potential the parameter $\varrho_{b}$ can uniquely be expressed
in terms of $E$ and vice versa.

In Section 5, we developed a general new method for constructing the approximate
instanton solutions for arbitrary potentials, which generalizes the well
known method of the thin-wall approximation \cite{Coleman} and actually allows 
to derive solutions outside of the validity of the thin-wall approximation. In the
leading order, the size of the instanton, its on-shell action and the potential
within the quantum core are presented either in formulae (\ref{55}) or (\ref{58}),
depending on $E$. These characteristics of the false vacuum decay are explicitly 
expressed in terms of the parameters that characterize the potential, namely its hight, 
the position of the false vacuum, and the coupling constants.

We have applied these formulas for the concave unbounded potentials
and derived in a very simple way the parametric dependence of the characteristics
of the instantons for a broad class of potentials which drop
as $-\varphi^{n}$ with $n\geq4$. It has been shown that in the
case of the quartic potential, for which there exists an exact solution
\cite{MRS2}, we have correctly reproduced, up to the numerical coefficients
of the order of unity, all the limiting cases elaborated in \cite{MRS2} as a
result of rather lengthy and tedious calculations. Although the numerical
coefficients can be quite large, as for example in the action $\frac{8\,\pi^{2}}{3\,\lambda_{-}}$,
the accuracy of our method does not allow us to estimate them reliably.
Therefore, we have focused on the parametric dependences of the solutions
and ignored the numerical factors, which can easily be computed
numerically for a given class of potentials by considering only a
single concrete set of parameters. Moreover, our method has allowed
us to obtain the dependence on the parameters even in those
cases for which the analytical solutions are not known, i.e., such as for
$-\varphi^{6}$ and other potentials. Our proposed method is applicable
to obtain the results in the leading order practically for arbitrary
potentials. The key element of this method is the ``master'' figure,
in which the dependence of the magnitude of the derivative of the
potential is compared with the value of the friction term as a function
of $\varphi$ calculated under the assumption that this term dominates
over the potential term in equation (\ref{5a-1}). In this way
we can find out which term in equation (\ref{5a-1}) can be neglected,
which in turn considerably simplifies the approximate solution in the corresponding
range of $\varphi$. We have shown how to apply our method
to the linear potential considered in the paper \cite{MRS1} and have
reproduced its main results in a few lines.

\bigskip{}

\textbf{Acknowledgments}

\bigskip{}

The work of V. M. was supported by the Germany Excellence Strategy---EXC-2111---Grant
No. 39081486.

The work of A. S. was supported in part by RFBR grant No. 20-02-00411.

\bigskip{}


\begin{thebibliography}{1}
\bibitem{Coleman} S.~R.~Coleman, \textit{The Fate of the False
Vacuum. Semiclassical Theory}, Phys.\ Rev.\ D \textbf{15} (1977)
2929 Erratum: {[}Phys.\ Rev.\ D \textbf{16} (1977) 1248{]}. doi:10.1103/PhysRevD.15.2929,
10.1103/PhysRevD.16.1248.

\bibitem{CGM} S.~R.~Coleman, V.~Glaser, and A.~Martin, \textit{\ Action
Minima Among Solutions to a Class of Euclidean Scalar Field Equations},
Commun.\ Math.\ Phys.\ \textbf{58} (1978) 211. doi:10.1007/BF01609421

\bibitem{MRS2} V. F. Mukhanov, E. Rabinovici, A. S. Sorin, \textit{Quantum
Fluctuations and New Instantons II: Quartic Unbounded Potential},
Fortsch.\ Phys.\ \textbf{69} (2021) 020101 doi:10.1002/prop.202000101
{[}arXiv:2009.12444 {[}hep-th{]}{]}.

\bibitem{MRS1} V. F. Mukhanov, E. Rabinovici, A. S. Sorin, \textit{Quantum
Fluctuations and New Instantons I: Linear Unbounded Potential}, Fortsch.\ Phys.\ \textbf{69}
(2021) 020100 doi:10.1002/prop.202000100 {[}arXiv:2009.12445 {[}hep-th{]}{]}.

\bibitem{Mukhanov} V. Mukhanov, \textit{Physical Foundations of Cosmology},
Cambridge University Press (2005).

\bibitem{MW} V. Mukhanov, S. Winitzki, \textit{Introduction to Quantum
Effects in Gravity}, Cambridge University Press (2007).

\bibitem{MS} V. F. Mukhanov, A. S. Sorin, \textit{On the Existence of the Coleman Instantons},
  arXiv:2104.12661 [hep-th].

\end{thebibliography}
\end{document}